CHARGE-GENERATED CHANGES IN GRAPHITE STRUCTURES

I. Pócsik[*], M. Veres, M. Füle, S. Tóth, M. Koós

Research Institute for Solid State Physics and Optics,

H-1525 Budapest, P.O. Box 49, Hungary

**Summary**

It will be shown that a negatively charged carbon atom in a graphite sheet has an electron structure that is unusual in carbons, viz. the lone pair of electrons. This atom, similarly to the positively charged ones, is no longer able to participate in the $\pi$ bond network: both pop out from the sheet because of their elongated single bonds and tetrahedral structure. The lone-pair structure localizes the extra electron. Forbidden gaps open above and below the Fermi level as a consequence of the destroyed $\pi$ bonds. This model offers an explanation for many 'anomalous' experimental data.

[*]Corresponding author, E-mail address: pocsik@szfki.hu

The splendid variability of the chemical bonds of carbon – such as the simple covalent bonds in diamond, the double bonds in benzene and graphite, and the triple bond in acetylene - was successfully explained 75 years ago by the elegant *sp* hybridization theory of Pauling[1-3]. The hybridized states form covalent $\sigma$ bonds with their neighbours, and the remaining non-hybridized *p* states form additional $\pi$ bonds with these, already bonded, neighbours. Concomitantly with the passage of time, the amount of contradictory data has also increased as a result of the intensive investigation of carbon structures. Possibly the most surprising is the disappearance[4-7] of the robust graphitic Raman spectrum[8-9] of the charged carbon electrode of the supercapacitor[10]. There are some other aspects relating to the electron structure of graphite intercalation compounds[11-13] that also need some alternative explanation.

A graphite sheet is generally accepted as an ideally planar structure by virtue of the non-hybridized p electron on each carbon atom. On the other hand, experimental data are available suggesting that this is not always the case. Quite a while ago Chow and Zabel[14] concluded from X-ray data that in certain cases the carbon layer in intercalated graphite should be buckled. Raman analysis [15] showed that the carbon atoms are not co-planar. It was recently demonstrated[16] that a graphite sheet shows corrugation, in its hexagonal manner. The sliding friction of graphite sheets on each other depends on the relative angle between the orientations of their crystalline axes. When these crystalline axes were parallel, the friction was 16 times greater, than in the twisted case.

It is commonly believed that by intercalation the removed compensating charges of the stored ions are stored on the graphite sheets thereby increasing the graphitic character of the compound. There are some ARPES measurements[17-20] – still unexplained – showing that this is probably not correct. The electron density in caesium intercalated graphite showed an around 3 eV wide forbidden gap below the Fermi level. This strongly contradicts our present knowledge since we will show that the plain graphite sheet is a finely balanced neutral structure that will be damaged by any deposited charge.

Negatively charged carbon in a graphite sheet shows a new configuration: the non-bonding pair, or 'lone pair' of electrons, that does not take part in chemical bonds, but occupies a tetrahedral segment of the space around the atom. The extra electron occupies the second site of the half filled, non-hybridized *p* state. This occupation transforms the *p* state into the lone-pair state thereby creating a nitrogen analogous $C^-$ electron structure - not mentioned in Pauling's list[1-3] of carbon. What we have is a three-coordinated, non-planar, tetrahedral structure of carbon atom, which pops out from the graphite sheet.

A further consequence is that the deposited charge destroys the former π bond of this $C^-$ atom, and its $C^o$ former partner. The bond type is reduced from double to single, and its length increases from around 141 pm in the graphite sheet to around 154 pm in diamond. This length increase causes these $C^-$ and $C^o$ atoms to pop out from the graphite sheet in an anti-symmetrical bi-pyramidal structure (see Fig. 1), contributing to an increase in the mean bond length reported by Pietronero and Strässler[21].

By charging or discharging a graphite sheet, no change takes place in the population of the π states, but their density changes: π states (both π and $\pi^*$) disappear. Ongoing charging/discharging leads to a gap opening in the inter-crossing region of the π and $\pi^*$ bands, as can be seen in Fig. 2.

The lone-pairs usually create energy levels around the inter-crossing point, so does the neutral $C^o$, or the $C^+$ atoms. The whole process is illustrated in Fig. 3. The band in the middle is narrow, created by the weakly interacting $C^-$, $C^o$ and $C^+$ carbons.

The displacement of charged atoms destroys the planarity of the sheet, and increases its corrugated character. This explains the measured Raman data[15] and the relative angle dependence of the friction of the graphite sheets[16].

The collapsed Raman spectra of charged ultra-capacitor electrode[4-6] can be explained by the increasing number of *sp³* carbons in the graphite sheet. The $E_{2g2}$ mode[13] is an 'in plane' vibration mode of the crystal sheet of graphite, but corrugation destroys the plane along which this vibration could propagate, thus the density of this vibration state decreases[4-6].

Direct proof of the forbidden gap opening is given by angle resolved photoemission spectra[17-20] measured on intercalated graphite. The π electron system dominates the electron spectrum of the graphite in the upper ~8 eV range, but the band intensity drops to zero in the upper ~3 eV range proving, that π bonds disappeared from this energy range. A sharp electron density peak at the 'Fermi energy' shows the increasing $C^-$ and $C^o$ concentration.

Graphite, generally accepted as being ideally flat, is the archetype of layered structures. Charges deposited on a graphitic sheet break up π bonds thereby creating three-coordinated carbons with tetrahedral electron structure, which seems to be the most characteristic and frequent defect in the graphitic sheet, transforming the flat sheet to a corrugated one. The broken π bonds are missing from the density of electron states, and a gap opens above and below the Fermi energy.

**Acknowledgements**

This work was supported by NATO; the Hungarian National Research and Development Program; and by the Hungarian Scientific Research Fund.

FIGURE CAPTIONS

Fig. 1. Displacement of $C^-$ and $C^o$ atoms from their position in the graphite sheet. Dashed lines show the original positions and bonds, arrows represent displacements.

Fig. 2. The $\pi$ and $\pi^*$ electron energy dispersion in the graphite sheet along highly symmetrical lines in the Brillouin zone of graphite, after Kresse et al.[22]. The thin line represents the pure single crystal, the thick one the electrically charged one, when gap ($E_g$) opened as a consequence of the broken $\pi$ bonds.

Fig. 3. Gap opening (shown by arrows) in the electron density of state of ongoing charged graphite.

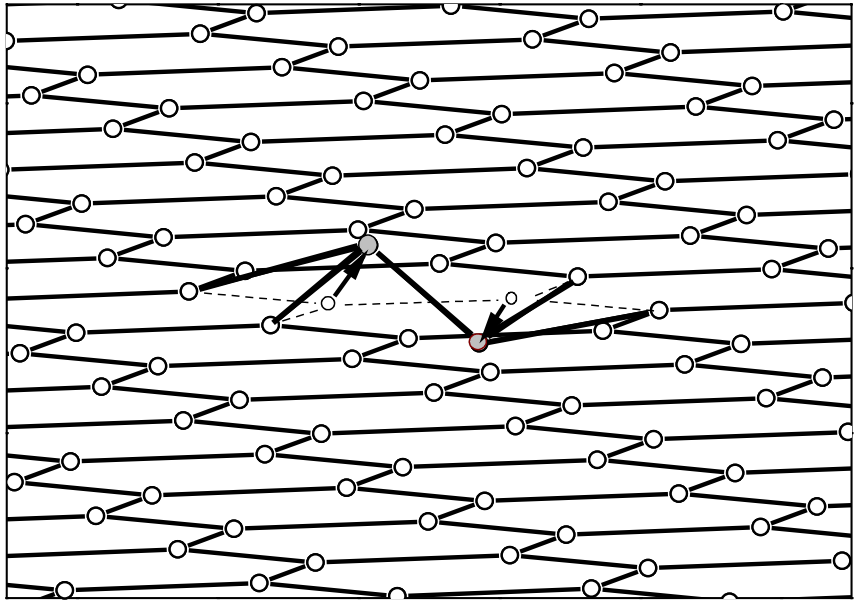

Fig. 1.

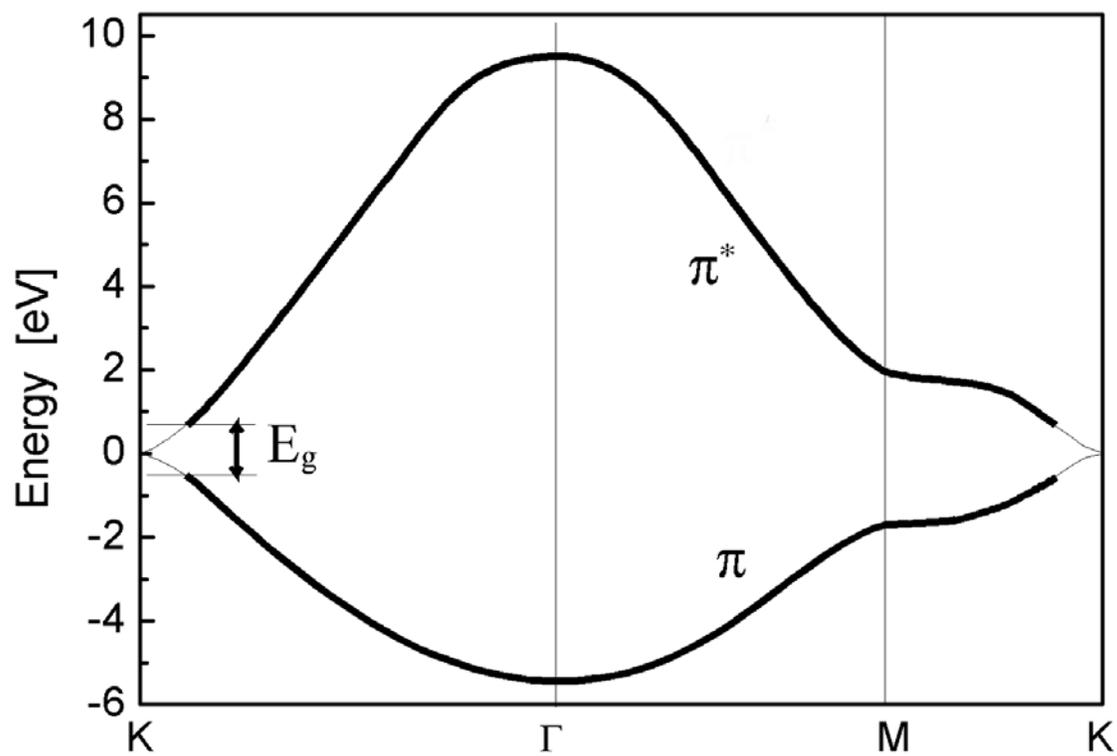

Fig. 2.

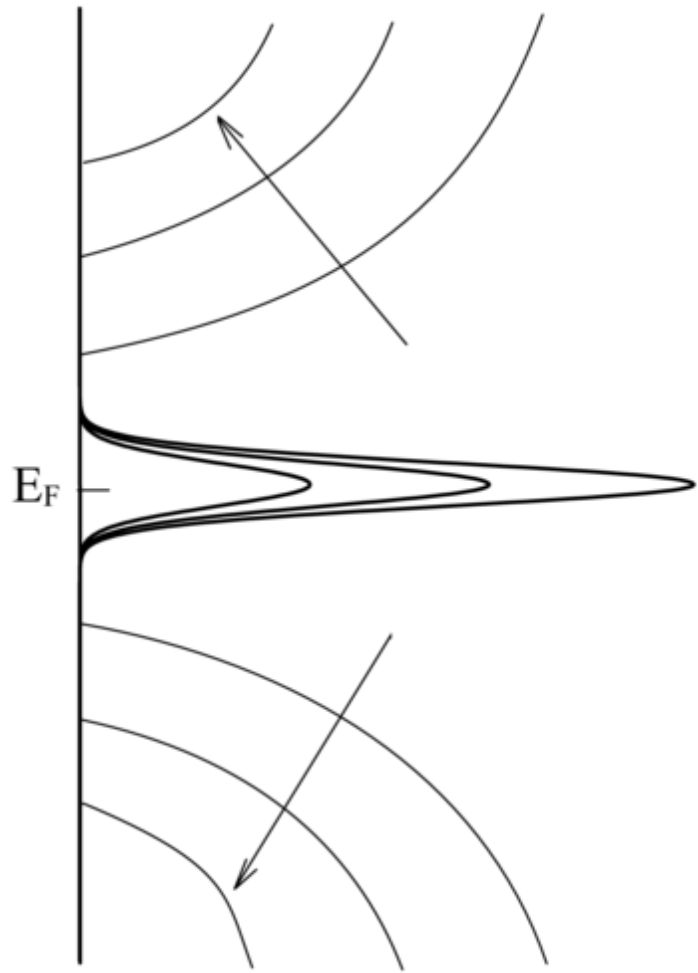

Fig. 3.